\documentclass[
    ,final            
  ]
  {aipproc}

\layoutstyle{6x9}

\usepackage{amsfonts}
\newcommand{\angstrom}{\mbox{\normalfont\AA}}

\begin{document}

\title{Observing scale-invariance in non-critical dynamical systems}


\author{Claudius Gros}{
}

\author{Dimitrije Markovi\'c}{
  address={Institute for Theoretical Physics, 
           Goethe University Frankfurt,
           60438 Frankfurt a.M., Germany}
}

\begin{abstract}
Recent observation for scale invariant neural avalanches 
in the brain have been discussed in details in the scientific 
literature. We point out, that these results do not necessarily 
imply that the properties of the underlying neural dynamics 
are also scale invariant. The reason for this discrepancy lies 
in the fact that the sampling statistics of
observations and experiments is generically biased by the 
size of the basins of attraction of the processes to be studied.
One has hence to precisely define what one means with
statements like `the brain is critical'.

We recapitulate the notion of criticality, as originally
introduced in statistical physics for second order phase 
transitions, turning then to the discussion of critical
dynamical systems. We elucidate in detail the difference 
between a 'critical system', viz a system on the verge of
a phase transition, and a 'critical state', viz state with
scale-invariant correlations, stressing the fact that the
notion of universality is linked to critical states.

We then discuss rigorous results for two classes of critical
dynamical systems, the Kauffman net and a vertex routing model,
which both have non-critical states. However, an external
observer that samples randomly the phase space of these two
critical models, would find scale invariance. We denote this
phenomenon as 'observational criticality' and discuss its
relevance for the response properties of critical dynamical
systems.

\end{abstract}

\maketitle


\section{Introduction}

The notion of criticality stems from statistical
mechanics and is fundamentally related to the deeply
routed concept of universality \cite{kadanoff1990scaling, 
stanley1999scaling}. As critical equilibrium systems show scale 
invariance it is natural to assume that the same would hold for 
critical non-equilibrium systems \cite{hinrichsen2000non, lubeck2004universal}. 
The situation is however substantially more complex for 
classical dynamical systems far from equilibrium and the subject 
of our deliberations. The discussion will revolve around 
three central concepts.

\begin{description}

\item{\sc critical system} \ A system is denoted critical when
  being located right on the transition point of a second order
  phase transition \cite{stauffer1972universality, balescu1975equilibrium}.

\item{\sc critical state} \ The state of a thermodynamic or 
  dynamical system is denoted critical when exhibiting scale
  invariance \cite{stauffer1972universality, kadanoff1989scaling}. 
  Critical thermodynamic systems dispose always of a critical state, 
  critical dynamical systems not necessarily.

\item{\sc observational criticality} \ The experimental observation
  of a dynamical system generically involves a stochastic sampling
  of its phase space. Scale invariance may be observed for a
  critical dynamical system which does not dispose of a critical
  state \cite{sornette2002mechanism, newman2005power, sornette2007probability}.

  This dichotomy is caused by the difference between mean and
  typical properties. It turns out that for critical dynamical systems
  the scaling behavior of the typical attractor may differ 
  qualitatively from the scaling of the mean attractor, as 
  defined by randomly sampling a phase space.

\end{description}

We will start by recapitulating the central notions of
the theory of critical thermodynamic systems, stressing
the fact that the scale invariance, which is observed in 
this case, is deeply intertwined with the concept of 
universality. We will then discuss two examples of critical 
dynamical systems for which the scaling behavior at 
criticality is, at least in parts, exactly known.

\begin{figure}[t]
  \includegraphics[width=0.7\textwidth]{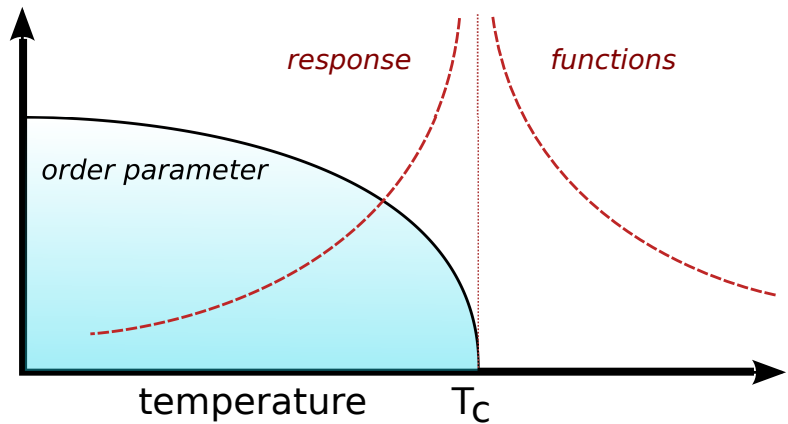}
  \caption{Illustration of a second order phase transition.
           The low-temperature phase is characterized by an order
           parameter which drops continuously to zero at the
           critical temperature $T_c$. The system becomes increasingly
           susceptible to perturbations coupling to the order parameter close
           to the transition point, the respective response functions
           diverge algebraically.}
\label{fig:responseFunctions}
\end{figure}

\section{Criticality in statistical physics}

In statistical physics a phase transition is termed a
second order phase transition when the ordering process 
starts continuously at the critical temperature $T_c$, 
when lowering the temperature $T$ of the system, 
compare Fig.~\ref{fig:responseFunctions}.
Otherwise, when the low-temperature state discontinuously
appears, one speaks of a transition of first order. The
theory of critical phenomena deals with second order
phase transitions \cite{yeomans1992statistical}.

\paragraph{Scaling towards criticality}

For a second order phase transition there are precursors
of the impending transitions, which can be measured 
experimentally using appropriate probes. For example, 
applying an external magnetic field to a ferromagnetic 
system will lead to a strong response, in terms of the 
induced magnetization, close to the transition. In general 
this response will diverge as
\begin{equation}
\sim\, \frac{1}{|T-T_c|^\gamma}~,
\label{eq_powlaw_T_T_c}
\end{equation}
where $\gamma>0$ is the critical exponent
\footnote{Critical exponents may differ for $T<T_c$ and
$T>T_c$}. Power-laws like Eq.~(\ref{eq_powlaw_T_T_c}) are denoted
scale invariant, as they do not change their functional
form when rescaling the argument via
$|T-T_c|\to c|T-T_c|$, where $c>0$ is an arbitrary
scaling factor.

\paragraph{Critical state}

At criticality, $T=T_c$, the thermodynamic state is very special,
its correlation function being scale invariant both in the
spatial and the temporal domain. For a magnetic system,
with moments $S({\bf x})$ at $\bf x$, the equal time
correlation function 
$$
D({\bf r})\ \equiv\ 
D({\bf x}-{\bf y})\ =\ 
\big\langle S({\bf x})
            S({\bf y}) \big\rangle
\,-\, \langle S\rangle^2
$$
obeys the scaling relations
\begin{equation}
D({\bf r})\ \propto\ \left\{
\begin{array}{rcl}
\mbox{e}^{-r/\xi} && T\ne T_c \\
r^{-\alpha} && T=T_c
\end{array}
\right.,
\qquad\quad \xi\propto \frac{1}{|T-T_c|^z}~,
\label{eq_correlation-function}
\end{equation}
with $\xi$ being termed the correlation length
and $z$ the critical dynamical exponent 
\cite{wang1997universality, hohenberg1977theory}.

\paragraph{Absence of microscopic length scales}
The scaling of the correlation function
(\ref{eq_correlation-function}) is very intriguing,
since it implies that all microscopic scales
(length, time, energy, {\it etc.}) become irrelevant 
at criticality. As an example consider the Schr\"odinger
equation
$$
i\hbar {\partial \Psi(t,{\bf r})\over \partial t} \ =\ 
-E_R\left(a_0^2\Delta \,+\, {2a_0\over |{\bf r} |}
           \right)\, \Psi(t,{\bf r})
\qquad \quad E_R={me^4\over 2\hbar^2}, \qquad \quad
a_0={\hbar^2\over me^2}
$$
which determines the properties of most matter we know.
The Schr\"odinger equation contains two scales,
the Rydberg energy $E_R=13.6\,\mathrm{eV}$, which determines
the energy level spacing, and the Bohr radius $a_0=0.53 \angstrom$,
which determines the extension of the atoms. Any
Hamiltonian known is characterized by corresponding scales,
but these become irrelevant at criticality and do not
determine the magnitude of the critical exponents.

\paragraph{Universality}
The symmetry of the high-temperature phase is broken 
at a second order phase transition. For example, in a 
magnetic systems with classical moments, these magnetic 
moments point in any direction for $T>T_c$, the symmetry 
of the high temperature phase is O(3), the symmetry
group of the sphere. In the low-temperature phase the
magnetic moments point however predominantly into 
a specific direction, breaking spontaneously 
the O(3) symmetry of the order parameter.

A central result of the modern theory of phase transitions
is now that the critical exponents are determined solely 
by two factors: the dimensionality of the system and the 
symmetry of the order parameter. This relation is termed 
`universality' as it allows to classify second order phase 
transitions into a relatively small number of distinct 
classes \cite{stauffer1972universality,kadanoff1990scaling,stanley1999scaling}. 
Results obtained using a given microscopic model are valid 
for all models within the same universality class. 
Universality is the core to our understanding of 
second-order phase transition, the scale invariance 
of the critical state being a manifestation of it.

\begin{figure}[t]
  \includegraphics[width=0.6\textwidth]{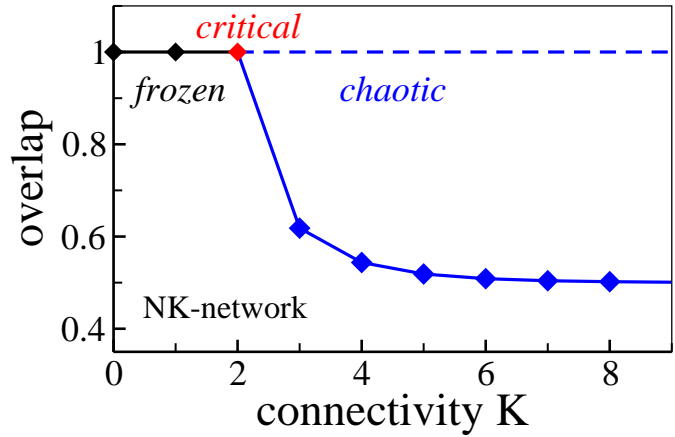}
  \caption{The evolution of the order parameter of the NK-network.
           Shown is the overlap, as given by Eq.~(\ref{eq_overlapp}),
           in the long-time limit, of two initially close trajectories.
           In the frozen state the overlap becomes maximal, since the 
           two trajectories flow into the same attractor. In the chaotic 
           state two initially close states diverge, the Lyapunov exponent
           is positive.
           }
\label{fig:cycle_kauffmanNetOrderParameter}
\end{figure}

\section{Boolean networks}

In equilibrium thermodynamics one studies 
systems in the thermodynamic limit where the number of 
components $N$ becomes infinitely large, $N \to \infty$.
Phase transitions hence take place, in statistical physics,
in systems made-up of many similar units. We consider
here an equivalent setting for non-equilibrium phase
transitions. A dynamical system can be described as a set 
of $N$ differential equations,
\begin{equation}
\frac{d}{dt}\, x_i(t) \ =\ f_i(x_1,..,x_N;\eta),
\quad\qquad i=1,..,N~,
\label{eq_dyanamical_system}
\end{equation}
where $f_i$ determines the time evolution of the 
dynamical variables $x_i(t)$ which are related to 
each of the system's elements. Here $\eta$ denotes 
a generic control parameter. Random Boolean
networks are defined by three specifications 
\cite{gros2008complex}.

\begin{description}

\item{\sc Boolean variables} \ The variables $x_i\in\{0,1\}$
  are Boolean and the time $t=0,1,2,\dots$ discrete.

\item{\sc random coupling functions} \ The coupling functions
  are Boolean, $f_i\in\{0,1\}$, and selected randomly.

\item{\sc connectivity} \ The coupling functions are determined
  by only a subset of $K$ randomly selected controlling elements and
  not by all $N$ Boolean variables. Hence the term `Boolean network'.
  The control parameter $K$ is denoted connectivity.
\end{description}

Random Boolean networks are also termed $NK$- or Kauffman 
nets \cite{kauffman1969metabolic}.
They show a phase transition for connectivity $K=2$, being
regular for $Z<2$ and chaotic for $Z>2$ \cite{socolar2003scaling}. 
\begin{center}
\setlength{\tabcolsep}{5pt}
\renewcommand{\arraystretch}{1}
\setlength\arrayrulewidth{1pt} 
\begin{tabular}{r|c|c}
 $K<2$ & $K=2$ & $K>2$ \\ \hline
frozen & critical & chaotic 
\end{tabular}
\end{center}
The order parameter is given by the overlap
\begin{equation}
\lim_{t\to\infty} \Big(1-||{\bf y}-{\bf x}||\Big)
\label{eq_overlapp}
\end{equation}
of two initially close trajectories ${\bf x}(t)$ and 
${\bf y}(t)$, where $||..||$ denotes the Manhattan 
distance, that is, the sum of the absolute differences 
of coordinates of ${\bf x}$ and ${\bf y}$. In the 
frozen phase the overlap is maximal, since
close-by trajectories will end up in the same attractor,
see Fig.~\ref{fig:cycle_kauffmanNetOrderParameter}.
The dynamics becomes chaotic however for $Z>2$, and
two trajectories diverge, with their mutual overlap 
decreasing.

\begin{figure}[t]
  \includegraphics[width=0.5\textwidth]{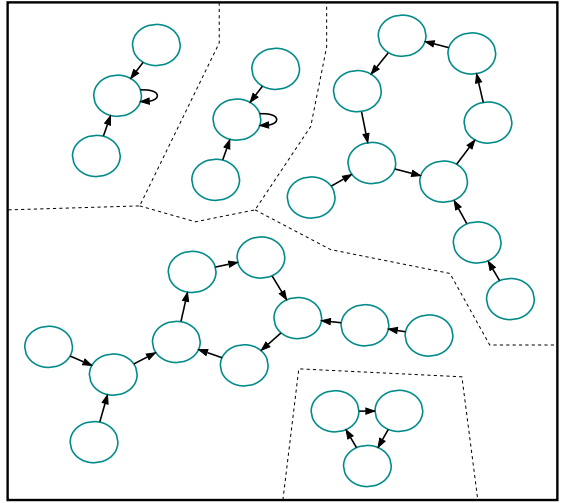}
 \caption{There are many cyclic attractors in the phase space
           of boolean networks and routing models. Each attractor
           comes with its distinct basin of attraction, which
           is made up of the cycle itself together with all
           points of phase space flowing into the attractor.
           }
\label{fig:cycle_basinsAtraction}
\end{figure}

\paragraph{Attractors and cycles}

The time evolution of any dynamical network with finite
phase space, which is $\Omega=2^N$ for the $NK$ net, is
determined by the number and the size of its cyclic 
attractors. The Kauffman net is critical for $Z=2$
and one may ask the question to which extend this 
criticality is reflected in the statistics of its 
attractors.

Any attractor comes with a respective basin of attraction, 
as illustrated in Fig.~\ref{fig:cycle_basinsAtraction},
defined as the set of all points in phase space
flowing into the attractor.
In the ordered phase a small number of attractors with
large basins of attraction dominates phase space and
the dynamics is hence very stable, nearby trajectories 
converge. In the chaotic phase, for $Z>2$, the number of
attractors is however very large and the size of 
their respective basins of attraction correspondingly
smaller. Nearby trajectories tend to diverge, being 
attracted by different cycles.

\paragraph{Finite-size scaling}

To calculate the properties of a dynamical or thermodynamic
system  directly in the thermodynamic limit is most of the
time difficult or impossible. Alternatively one can 
evaluate the quantity of interest for finite systems size
$N$ and then extrapolate to large system size, a 
procedure denoted finite-size scaling. For scale invariant 
states, like the critical thermodynamic state, finite
size scaling involves power-laws. The reason is that there
are no length scales at criticality in statistical physics 
and power-laws are the only scale invariant relations.
Conversely we expect finite-size scaling to be algebraic
whenever the underlying state is critical, viz scale
invariant.

Initial numerical calculation for the $Z=2$ Kauffman net
did indeed find that the number of attractors, scales 
polynomial, like $\sqrt{N}$ \cite{kauffman1969metabolic}. 
The same scaling relation was also found for the mean cycle length. 
However it has recently been show rigorously,
that the number of attractors actually increases faster than
any power of $N$, viz super-polynomial \cite{socolar2003scaling,
mihaljev2006scaling}. The intrinsic state of the critical 
$Z=2$ Kauffman net is hence not scale invariant.

\begin{figure}[t]
\centerline{
\includegraphics[width=0.25\textwidth]{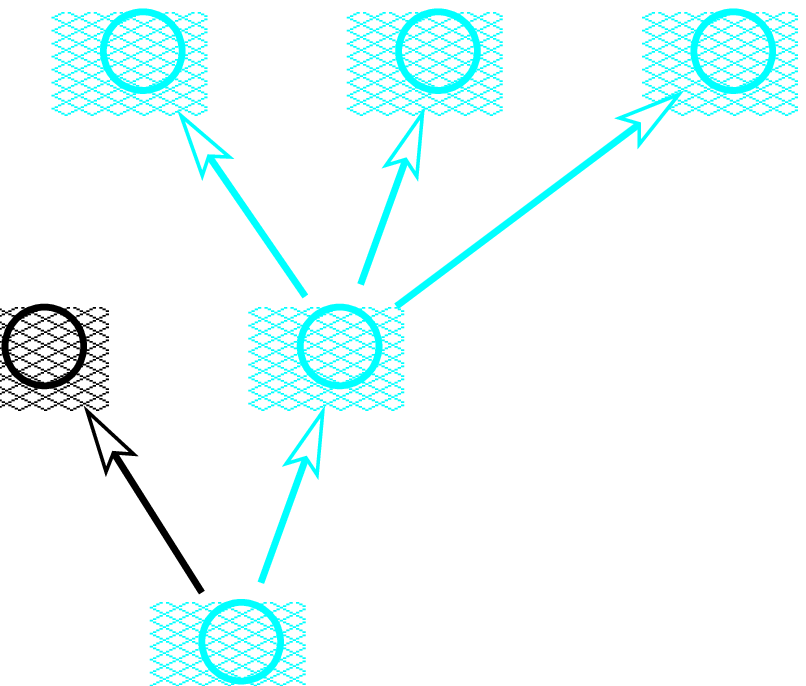}
\hspace{3ex}
\includegraphics[width=0.25\textwidth]{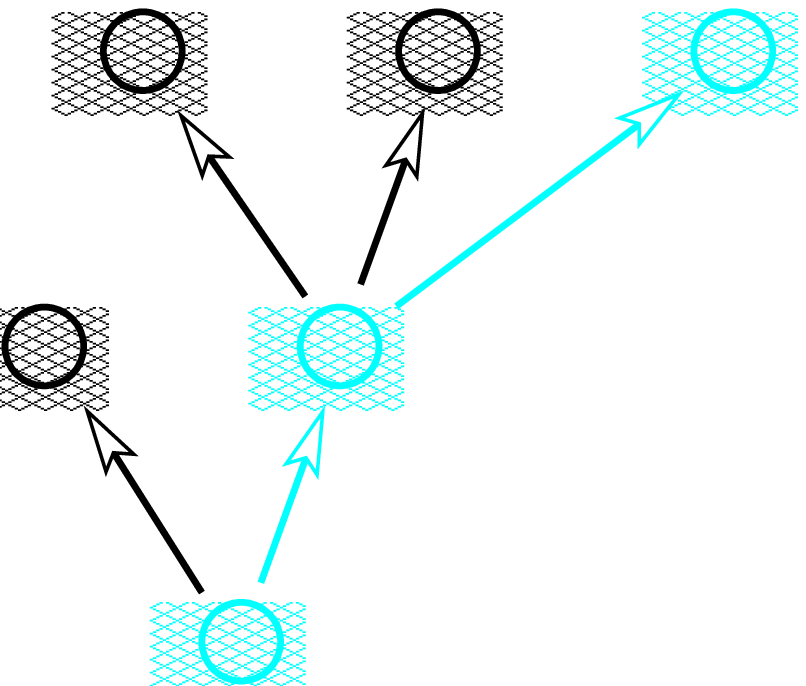}
\hspace{3ex}
\includegraphics[width=0.25\textwidth]{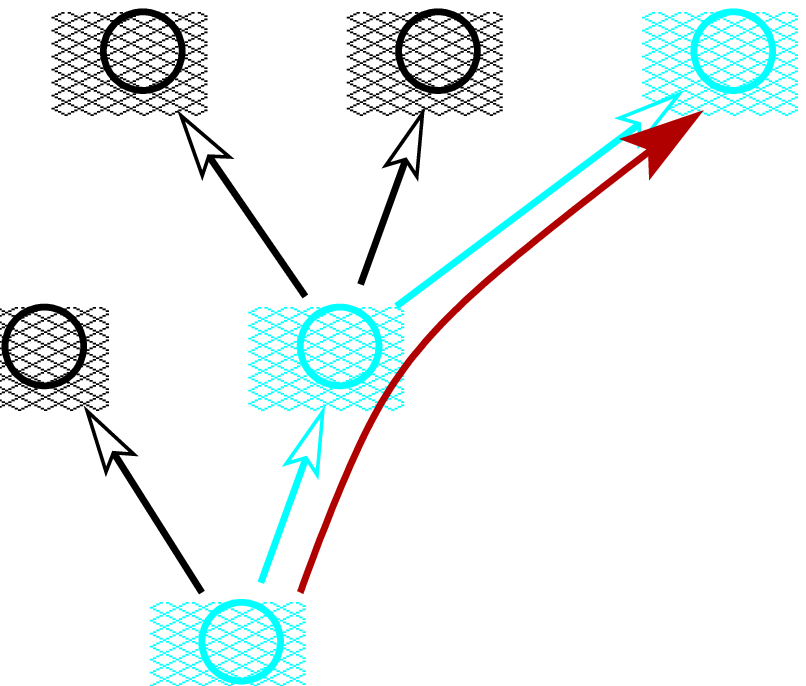}
           }
  \caption{Illustration of information spreading on networks.
           When information spreads diffusively (left), it may
           be passed on to any number of subsequent vertices.
           When information is conserved (center), the information
           can be considered as a package which can be passed on only
           to a single downstream site. Alternatively one can
           consider information routing (right), where an incoming
           package is routed to an outgoing link.}
\label{fig:informatinSpreading}
\end{figure}

\paragraph{Observational scale invariance}
The phase space $\Omega=2^N$ of the $NK$ network increases
exponentially with system size $N$. Numerical studies
have hence to resort to appropriate statistical sampling
of phase space. Actually, this is also what an experimental
observer would do when examining a dynamical system at
a random starting time. It may now be the case that a 
relatively small number of attractors dominate phase 
space and the results of a statistical sampling 
procedure, see Fig.~\ref{fig:cycle_basinsAtraction}.

In order to illustrate this scenario we discuss now a
fictional example. Let's assume that there are big attractors 
of the order of $\sqrt{N}$, each having on the average 
a basin of attraction of the size
$$
\sim\ \frac{\Omega}{\sqrt{N}} \ =\ \frac{2^N}{\sqrt{N}}~.
$$
There could be in addition a very large number of point
attractors, each having a basin of attraction of size one.
For example the number of point attractors could scale 
super-polynomial like 
$$
\sim\ 2^{\sqrt{N}}~.
$$
In this case their combined relative contribution 
$$
\sim\ \frac{2^{\sqrt{N}}}{\Omega} \ =\ 
\frac{2^{\sqrt{N}}}{2^N} \ =\ \frac{1}{2^{\sqrt{N}}}
$$
to phase space would still vanish in the thermodynamic 
limit $N\to\infty$. This is what happens for the $Z=2$ 
Kauffman net. The typical attractor is very small and 
not seen by a stochastic sampling procedure. A relatively small
number of big attractors with large basins of attraction 
dominate phase space and determine the statistics as sampled
by an external observer.

\begin{figure}[t]
\centerline{
\includegraphics[height=0.33\textwidth]{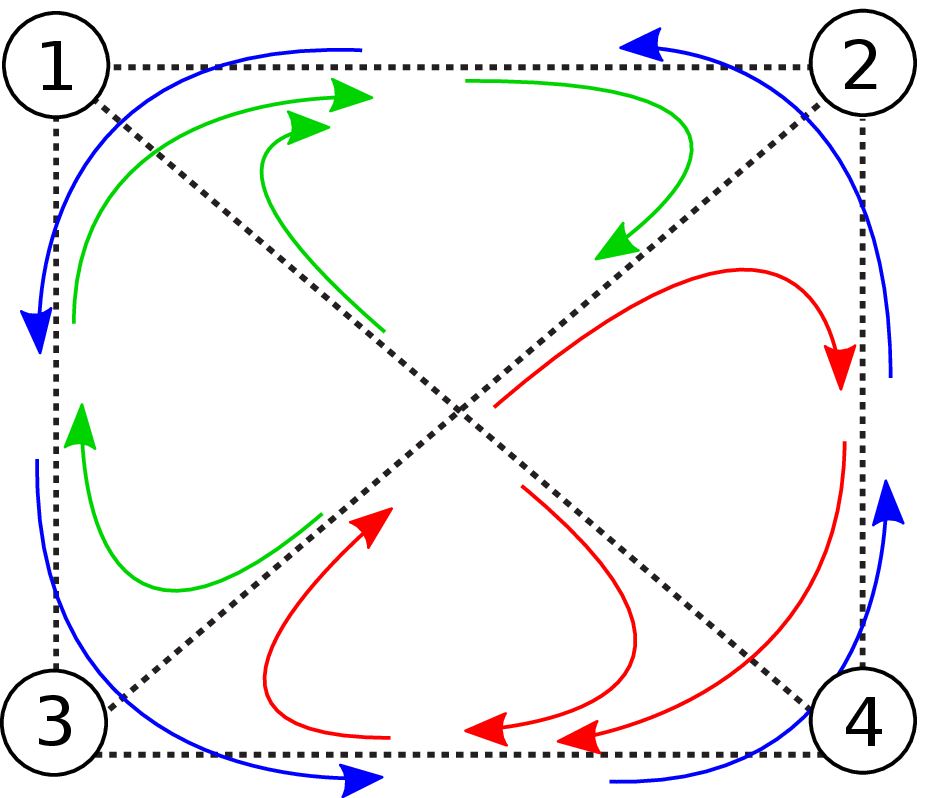}
\hspace{8ex}
\includegraphics[height=0.33\textwidth]{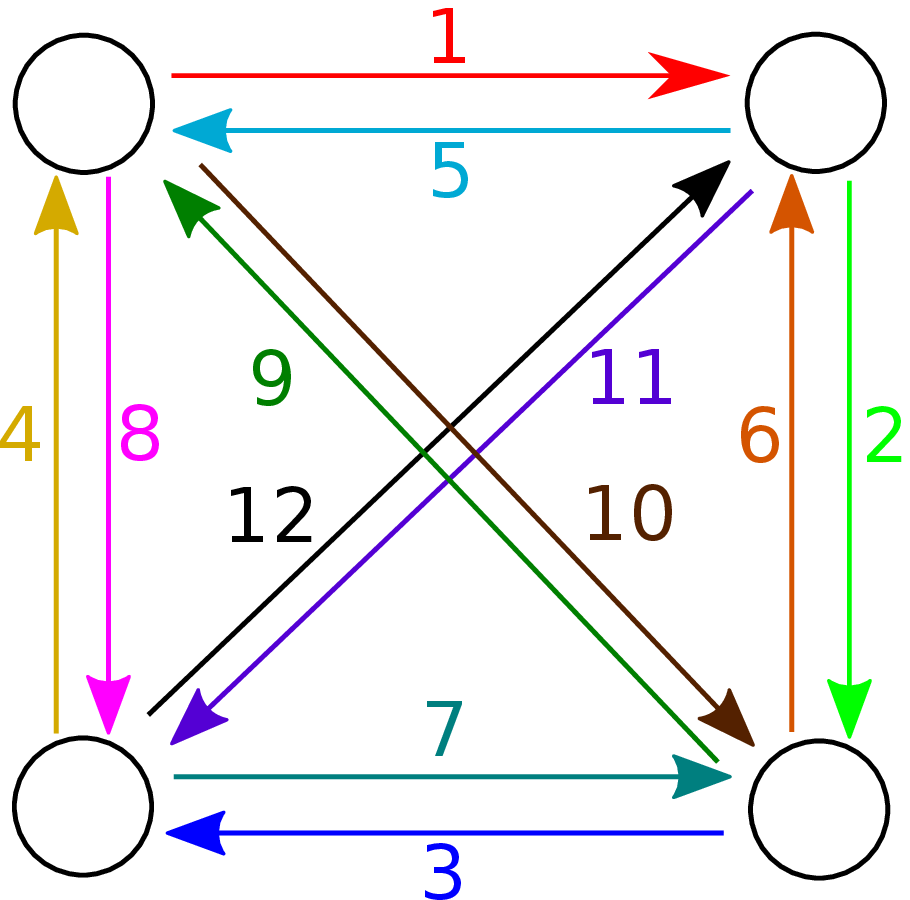}
           }
  \caption{Illustration of of a $N=4$ sites vertex routing model
           which has (left) three cyclic attractors. Note that more 
           than one cycle can pass through any given vertex, as
           the phase space (right) is made up by the collection of
           the $N(N-1)=12$ directed links.}
\label{fig:vertexRoutingModel}
\end{figure}

\section{Vertex routing models}

Criticality and conservation laws are intrinsically 
related. A branching process is critical, to give
an example, when the average number of offspring
is equal to the number of parents, that 
is, when average activity remains constant.
It is hence possible to construct critical dynamical
systems when incorporating a conservation
of activity levels. An example for this procedure are
vertex routing models \cite{markovic2009vertex}.

Information can spread diffusively or via routing processes,
see Fig.~\ref{fig:informatinSpreading}. For the later case
one considers information packages transmitted at every
vertex via randomly selected routing tables. The phase 
space is hence given by the collection of directed links,
the phase space volume $\Omega=N(N-1)$ scales algebraically.
More than one cycle can hence pass through a given vertex.
The number of cycles passing through a given model
can be viewed as a measure for information centrality
which has a non-trivial distribution in the thermodynamic
limit \cite{markovic2009vertex}.

\paragraph{Exact solution}
The routing dynamics can be mapped to a random walk in
configuration space, the collection of directed links,
and solved exactly \cite{kruskal1954expected, gros2008complex}. 
The number $\langle C_L\rangle(N)$ of cycles of length $L$ is 
given by
\begin{equation}
\langle C_L\rangle(N)\ =\ 
\frac{N((N-1)^{2})!}{L(N-1)^{2L-1}((N-1)^{2}+1-L)!}~,
\label{eq_VR_C_L_quenched}
\end{equation}
for fully connected graphs with $N$ vertices.
In addition to the exact expression 
(\ref{eq_VR_C_L_quenched}) for the intrinsic
cycle length distribution of the routing model, one
can also derive the distribution of cycle length
an observer would find when randomly sampling
phase space. In this case the probability to 
find a given cycle of length $L$ is weighted by
the size of its basin of attraction. The
resulting cycle length distribution is
\begin{equation}
\langle C_L\rangle(N) \ \propto\ 
\sum_{t=L}^{L_{max}}\,
\frac{((N-1)^{2})!}{(N-1)^{2t}((N-1)^{2}+1-t)!}~.
\label{eq_VR_C_L_on-the-fly}
\end{equation}
Algorithmically the difference between the
expressions (\ref{eq_VR_C_L_quenched}) and
(\ref{eq_VR_C_L_on-the-fly}) is equivalent 
to quenched deterministic and on-the-fly stochastic 
dynamics. Quenched dynamics is present when the routing 
tables are selected once at the start and then kept 
fixed, whereas for on-the-fly dynamics one randomly 
generates an entrance to a routing table `on the fly', 
viz only when needed.

\begin{figure}[t]
\centerline{
\includegraphics[height=0.3\textwidth,angle=0]{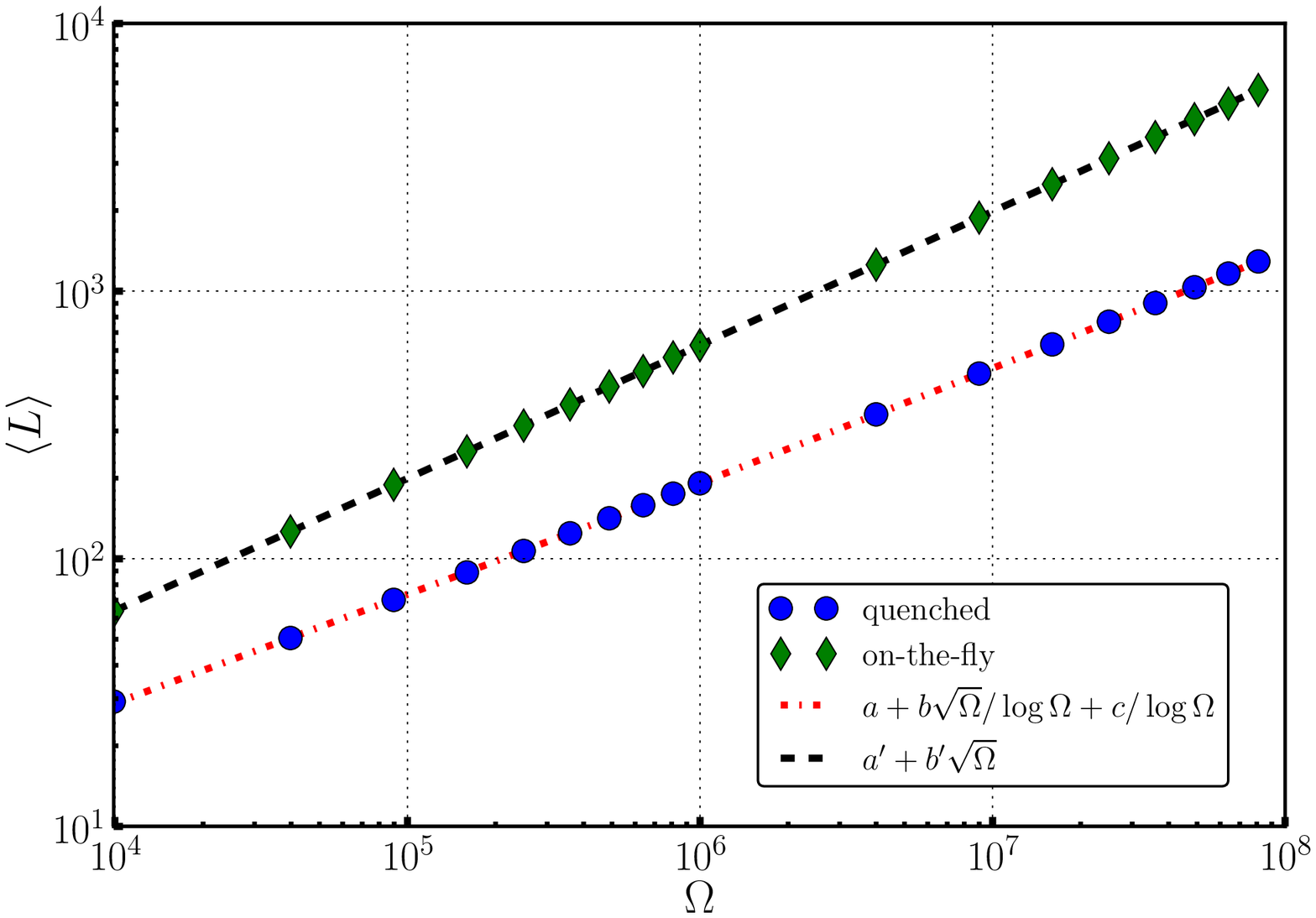}
\hspace{4ex}
\includegraphics[height=0.3\textwidth,angle=0]{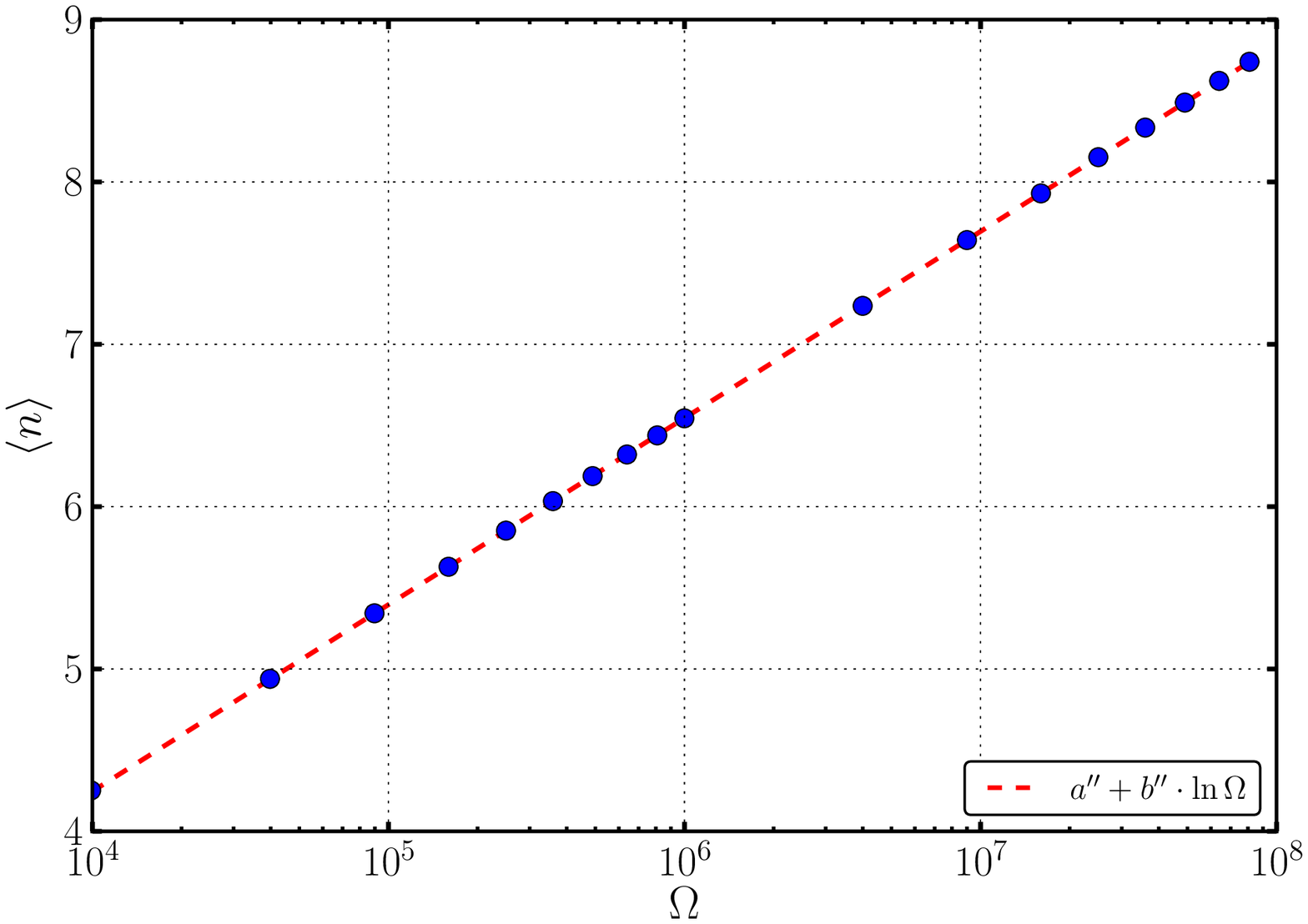}
           }
  \caption{Exact results for the vertex routing model. The
           mean cycle length (left) for both quenched and
           on-the-fly dynamics and the the mean cycle number
           (right), which can be evaluated only for
           quenched dynamics.}
\label{fig:vertexRoutingModel_results}
\end{figure}

\paragraph{Scaling of the vertex routing model}
One can evaluate the exact
expressions (\ref{eq_VR_C_L_quenched}) and
(\ref{eq_VR_C_L_on-the-fly}) for very large 
system size $N$, the results are shown in
Fig. \ref{fig:vertexRoutingModel_results},
respectively for the average cycle length
$\langle L\rangle$ and the overall number of cycles.
Only relative quantities can be evaluated with
on-the-fly dynamics and hence $\langle L\rangle$
but not the total number of cycles present.
The results are given in Table~\ref{table:VR-results},
where we have included also results for a modified
vertex routing model, a Markovian variant.
On-the-fly routing results in power-law scaling for
the average cycle length, in contrast to the exact
properties of the respective model, which contains
logarithmic corrections.

\begin{table}[t]
\begin{tabular}{|l|l|l|l|}
\cline{3-4}
\multicolumn{2}{c|}{}
 & \ quenched \ & \ on the fly \ \\
\hline \
\parbox[c]{0.13\textwidth}{vertex\newline routing}
 & \
\parbox[c]{0.25\textwidth}{ number of cycles\newline mean cycle length}
 & \
\parbox[c]{0.14\textwidth}{$\log(N)$\newline $N/\log(N)$}
 & \
\parbox[c]{0.1\textwidth}{--\newline $N$}
 \\
\hline
\parbox[c]{0.13\textwidth}{markovian\newline model}
 & \
\parbox[c]{0.25\textwidth}{number of cycles\newline mean cycle length}
 & \
\parbox[c]{0.14\textwidth}{$\log(N)$\newline $\sqrt{N}/\log(N)$}
 & \
\parbox[c]{0.1\textwidth}{--\newline $\sqrt{N}$}
 \\
\hline
\end{tabular}
\caption{The scaling behavior of the vertex routing model
         (first row) and of a modified routing model with
         nor routing memory (second row). Corrections
         $\sim\log(N)$ are present for quenched dynamics,
         viz for the intrinsic model behavior. An observer
         would however obey power-law scaling, as given by
         the on-the-fly dynamics, which can evaluate only
         relative quantities (and not the overall number of cycles).
        }
\label{table:VR-results}
\end{table}

\section{Discussion}

When probing a dynamical or thermodynamical system,
like the brain or a magnet, one needs to perturb the
system and measure the resulting response. The probing
protocol may be considered unbiased when the phase
space is probed homogeneously. If the dynamical system
being probed contains attractors, or attractor relics
\cite{gros2007neural,gros2009cognitive}, these will
dominate the statistics of the response. It may now
happen that properties of the attractors, like the
cycle length for the case of cyclic attractors, have a
highly non-trivial statistics in the sense, that the characterizing 
properties of the typical attractor differ qualitatively
from the average behavior probed by random sampling phase space.
In this the intrinsic or typical properties of the system 
differ from the one an observer would find when 
sampling phase space randomly.

We have argued in this study, that this situation does
indeed occur for critical dynamical systems, at least
for the classes of critical systems for which exact
results are known, Boolean networks and vertex routing models.
We believe that further investigation into this question
is warranted for additional classes of critical dynamical
systems, in order to examine the question whether power-law
scaling is independent, or conditional, on universality in
critical dynamical systems. This is an open issue. Here we
found that the intrinsic state of two critical dynamical 
systems is not scale invariant, a property typically associated
with universality in thermodynamics, but experimentally probing
the system stochastically would result in power-law scaling.


\begin{theacknowledgments}
  We thank the German Science Foundation (DFG) for
  financial support.
\end{theacknowledgments}

\bibliographystyle{aipproc}   
\bibliography{references}

\end{document}